\documentstyle[prl,twocolumn,aps]{revtex}

\begin{document}
\input epsf
\epsfclipon
\draft

\wideabs{

\title{Scaling and Crossovers in Diffusion Limited Aggregation}
\author{E. Somfai$^1$, L. M. Sander$^1$, and R. C. Ball$^2$} 
\address{
$^1$ Department of Physics, The University of Michigan, Ann Arbor MI 
48109-1120. USA\\
$^2$ Department of Physics, The University of Warwick, Coventry
CV4 7AL, UK}

\date{\today}

\maketitle

\begin{abstract}
We discuss the scaling of characteristic lengths in diffusion limited
aggregation (DLA) clusters in light of recent developments using
conformal maps. We are led to the conjecture that the apparently anomalous
scaling of lengths 
is due to one slow crossover.
This is supported by an analytical argument for the scaling of the
penetration depth of newly arrived random walkers, and by numerical
evidence on the Laurent coefficients which uniquely determine each
cluster. We find a single crossover exponent of $-0.3$ for all
the characteristic lengths in DLA. This gives a hint about the
structure of the renormalization group for this problem. 
\end{abstract}

\pacs{PACS numbers: 64.60.Ak, 61.43.Hv}

} 


Diffusion-limited aggregation (DLA) is a model for
growth that was introduced in 1981 by Witten and Sander \cite{WS} and
has been the subject of intensive scrutiny \cite{Mbook} ever
since. Remarkable progress has been made in numerical treatments of
the model and in applications to various physical
problems. Nevertheless, a fundamental understanding of the most
striking features of DLA clusters is still lacking.  A very promising
recent development is the method of iterated conformal maps
introduced by Hastings and Levitov \cite{HL} (HL) and pursued by
Davidovitch et al.  \cite{Benny}.  Using the method of conformal maps we
will try to
clear up one of the major difficulties in the field, namely the large
corrections to scaling which lead to slow crossovers. This discussion
has a very significant outcome: there has been serious doubt about
whether DLA clusters are really fractal, based on the 
existence of characteristic lengths apparently scaling {\it differently} from
the overall cluster radius. We show that these lengths cross over to scale in 
the same way as the radius, and suggest that all of the various
deviations from exact fractal scaling have the same source, with
the same correction to scaling exponent. There is, in our view,
no longer any reason to doubt that DLA clusters are fractal.

In the DLA model a nucleation center is fixed at the origin of
coordinates and random walkers are released from outside. When a
walker comes in contact with the cluster it sticks. The process
continues until $N$ walkers have attached; in modern work $N$'s of
$10^6$ are reasonably easy to attain.  Provided lattice anisotropy is avoided,
the clusters so produced seem
to be self-similar with fractal dimension $D \approx 1.71$ in two
dimensions and the probability of growth at a point on the surface
appears to be multifractal.

The conformal map method uses the {\it Laplacian growth}
\cite{WS83,NPW} version of DLA.  That is, we take the cluster to be a
grounded conductor in the $z$ plane, with a probability to grow at a
point on its surface proportional to the charge there: $ |\nabla V|$,
where $V$ is the potential with boundary conditions of unit flux at
infinity and $V=0$ on the surface.  If we construct a complex
potential such that $Re[\Psi(z)]=V$, and define $w=e^\Psi=Z^{-1}(z)$,
then $Z(w)$ is a conformal map which takes the exterior of the unit
circle in the $w$ plane to the exterior of the DLA cluster in the $z$
plane, and is linear, $Z \sim r_o w$, for large $|w|$.  Thus $V
\rightarrow \ln(|z|/r_o)$ at large $|z|$. The Laplace radius, $r_o$,
is the radius of the grounded disk with the same capacitance (with respect to
any distant reference point) as the
cluster \cite{Benny}.
The growth probability is $1/|Z'|$. Intervals $d\theta$ on the unit
circle in the $w$ plane correspond to intervals of arc length $ds$
with equal growth probability in the $z$ plane. 

To characterize the map we write:
\begin{equation}
\label{Laurent}
Z(w)  = r_o w +\sum_{k=1}A_k/w^k
\end{equation}
The Laurent coefficients, $A_k$, are a useful parameterization of the
map.  In what follows we will assume that the Laurent expansion of $Z$
has no constant term. This corresponds to moving the DLA cluster so
that the center of charge is at the origin.

In the HL method \cite{HL,Benny} the $A_k$ are produced directly.
However, this method employs certain approximations, and also numerically
difficult for large $N$ because it is of order $N^2$, and as a
practical matter is limited to $N \approx 10^4$.
We observe that there is a way around these problems.
Suppose we produce a DLA cluster by the conventional method
with random walkers (which is much faster) and then freeze it at some
$N$. Then by recording where $M$ random walkers would attach to the
cluster we have a set of points $z_m$.  These are at angle $\theta_m
\approx 2\pi (m/M)$ in the $w$ plane, since we are sampling the
charge, and equal increments of charge correspond to equal increments
of $\theta$.  The $A_k$ are the Fourier coefficients of the function
$z(\theta_m)$. We should note  that everything in this
paper results from the existence of the conformal map, 
and the behavior of the $A_k$, not on the
detailed method of generation proposed by HL. 

\begin{figure}
    \epsfxsize=3.4in
    \noindent\hfil\epsfbox{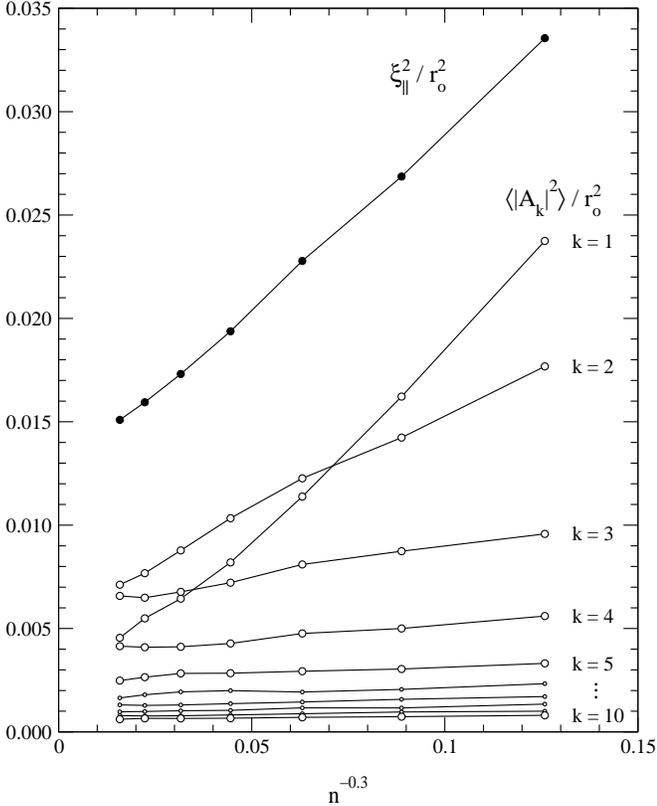}\hfil\bigskip\bigskip
    \caption{Crossover of the Laurent coefficients (open circles) and
    $\xi_\parallel^2$ (filled circles). $\xi_\parallel^2$ is approximated by
    the first 10 terms of the rapidly converging sum in Eq.~(\ref{penet_sum}).
    The size of the clusters range from $10^3$ to $10^6$ in steps of
    $\sqrt{10}$, and for each given size an ensemble of 1000 was taken.}
    \label{fig-crossover}
\end{figure}

Numerically we have found, using either the HL method \cite{Benny} or the
conventional method described above (cf. Fig.~\ref{fig-crossover}), that $r_o
\propto N^{1/D}$ and that for all but the first few $k$'s,
$\langle|A_k|^2\rangle \propto N^{2/D}$.
For $k\lesssim 4$ the $A_k$ appear to scale with a smaller power of $N$, a
behavior that we will interpret below as a crossover.

We will be concerned first with penetration depths of the random
walkers, i.e. how far from the origin they land.  This has been a
matter of considerable interest for some time because the width of the
distribution of this quantity (``the width of the growth zone'') seems
to increase more slowly with $N$ than the mean radius \cite{PR}.  This
is a disturbing observation since a real fractal should have no length
scale other than its overall size. Some authors \cite{MeakS} have
maintained that what was observed was a slow crossover and that
asymptotically all lengths scale together, but others \cite{Mandel}
have given evidence for an 'infinite drift scenario' in which DLA is
not a fractal at all in the asymptotic limit.  The structure of
conformal map theory gives us a very elegant way to discuss these
matters. 

\begin{figure}
    \epsfxsize=3.4in
    \noindent\hfil\epsfbox{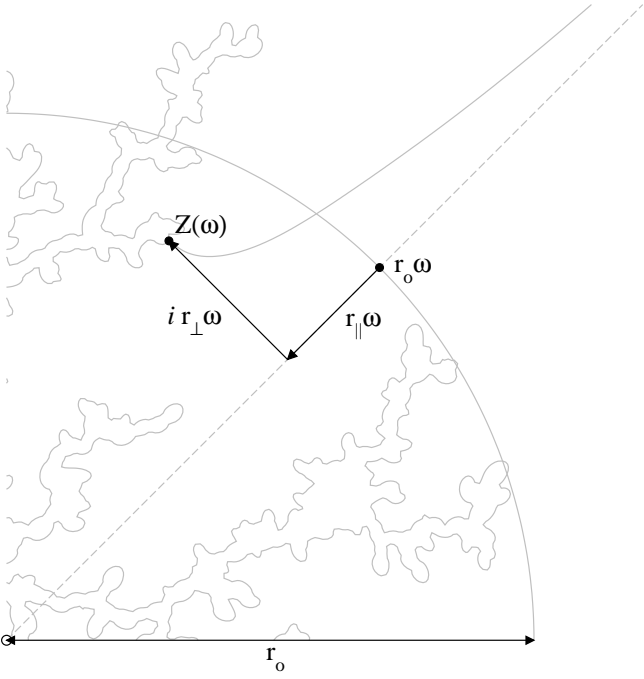}\hfil\bigskip\bigskip
    \caption{The radial displacement $r_\parallel$ and the transverse
    displacement $r_\perp$. The origin is at the center of charge.}
    \label{fig-diagram}
\end{figure}

We can define a penetration depth as follows: suppose we follow a 
field line from large $|z|$ to the surface of the cluster. The displacement
of the endpoint of the line from where it would terminate on the 
equivalent disk may be written as
$ (r_{\parallel} + i r_{\perp}) w \equiv Z(w) - r_0 w$,  where $r_{\parallel}$
is the radial displacement and $r_{\perp}$ is the transverse displacement, see
Figure~\ref{fig-diagram}.
The spread of $r_{\parallel}$ defines a penetration depth $\xi_{\parallel}$:
\begin{equation}
  \xi_{\parallel}^2 = \frac{1}{2\pi} \oint r_{\parallel}^2 d\theta .
\end{equation}
We also consider the analogous quantity $\xi_{\perp}$:
\begin{equation}
  \xi_{\perp}^2 = \frac{1}{2\pi} \oint r_{\perp}^2 d\theta .
  \label{xiperp}
\end{equation}
Using Eq.~(\ref{Laurent}) we have  
\begin{equation}
  \xi_{\parallel}^2 = 
  \xi_{\perp}^2 = \frac{1}{2} \sum_k |A_k|^2 .\label{penet_sum}
\end{equation} 
This exact result has a useful geometric interpretation: penetration of
a DLA cluster corresponds to landing inside one of the ``fjords''.
Clearly $r_{\perp}$ is the azimuthal deviation of a walker from
a straight path, and $r_{\parallel}$ the penetration. Since they
are the same, on average, the only way for DLA to have an
anomalous scaling for the penetration depth is for the 
angular width of the
fjords to be smaller and smaller as the cluster grows. 

Now we can discuss the
scaling of $\xi_{\parallel}$, the quantity of most interest, by
analyzing the $A_k$ and using Eq.~(\ref{penet_sum}).
We have remarked above that the numerical evidence is that the 
most of the $|A_k|$ 
scale as the radius. This is, in some sense obvious, since there
is an elementary formula that relates the area, $S_N$, of the 
image of the circle to the Laurent coefficients:
\begin{equation}
\pi r_o^2 = S_N  + \pi \sum_{k=1} k|A_k|^2
\label{area}
\end{equation}
The right hand side scales as $N^{2/D}$, but the area only as $N$.  
The leading behavior of the sum must also be $N^{2/D}$. The obvious
way for this occur is if each coefficient has this scaling, which would
lead to $\xi_{\parallel}^2  \propto  r_o^2$. 
We will now show that this expectation is correct. 

If we start with all the $A_k$ small, then in the course of growth,
they will not remain small. This is because Laplacian growth is
subject to the Mullins-Sekerka instability \cite{Mull}. In fact the
classic calculation of Ref. \cite{Mull} was concerned with exactly
this behavior: the Fourier coefficients of the wrinkling of the
surface grow exponentially, with the growth rate proportional to
$k$. The slowest growth is associated with the smallest $k$. However,
we are concerned here with the limiting behavior of $A_k$, far out of
the linear regime.

We can get some insight by making the following estimate for large $k$:
\begin{eqnarray}
\label{estimate}
|A_k|^2 & = & \frac{1}{4\pi^2} 
\int d\theta_1 d\theta_2 Z(e^{i\theta_1})Z^*(e^{i\theta_2})e^{ik(\theta_1
-\theta_2})
\nonumber \\
& = & \frac{1}{2\pi^2 k^2}\int dZ(e^{i\theta_1}) dZ^*(e^{i\theta_2})
  \cos (k[\theta_1 - \theta_2])  
\nonumber \\
 & \approx &  \frac{1}{4\pi^2 k^2} \sum_{\delta \theta = 1/k} |\Delta
 Z(e^{i\theta})|^2
\end{eqnarray}

The sum in the last equation can be evaluated in terms of the
multifractal spectrum, $\tau(q)$. This follows from the partition
function approach of Halsey, et al.  \cite{Hetal} where the spectrum
is defined by the implicit equation:
\begin{equation}
\sum_{m=1}^k (P_m)^q (|\Delta Z_m|/r_o)^{-\tau(q)} =1
\end{equation}
Here the surface of the cluster should be thought of as being divided
into $k$ boxes. The charge in box $m$ is $P_m$, and the box size is
$|\Delta Z|$ for that box. In our case the charges are all equal to
$1/k$, and we must put $\tau = -2$. Then we have:
\begin{eqnarray}
|A_k|^2 & \sim & (r_o^2/k^2) \sum [|\Delta Z|/r_o]^2
\nonumber \\
 & \sim & r_o^2 k^{\hat{q}-2}
\label{Ak}
\end{eqnarray}
where $\tau(\hat{q})=-2$ \cite{excuse}. We know that the function $\tau(q)$ is
increasing, and that $\tau(0) = -D$. Thus $\hat{q} < 0$ and we confirm that
the sum in Eq.~(\ref{area}) is dominated by its first few terms, and that each
scales as $r_o^2$.  Furthermore we have
$\xi_{\parallel}^2 = \frac{1}{2} \sum_k |A_k|^2 \sim r_o^2$ asymptotically,
and the sum converges rapidly.

Thus the asymptotic scaling behavior of the sum for the penetration length,
Eq.~(\ref{penet_sum}), is the same as that of its terms and the sum is dominated
by the
first few.  The crossover in the penetration depth must be
associated with the crossover of {\it the first few Laurent coefficients}
which, as we have seen, appear to behave differently from the
prediction of Eq.~(\ref{Ak}).  The first few coefficients 
(see Ref. \cite{Benny})
represent the quadrupole, octupole, etc. moment of the cluster. We can
guess that these have intrinsically slow dynamics (as in the linear regime)
and thus cross over more slowly than higher moments. Any quantity
associated with averages over the charge (the growth zone) should
share this crossover. In fact, we conjecture that there is one kind of
crossover of lengths with the same correction to scaling exponent.

We can verify this picture by numerical analysis of the penetration depth and
other moments of the charge. We computed the Laurent coefficients $r_o$ and
$A_1,\ldots,A_{10}$ for conventional DLA clusters using the method described
earlier. For each cluster the number of points $M$ sampling the harmonic
measure was equal to the size of the cluster, but at least $10^5$; this way
the numerical error in the computed coefficients is around 10\%, and the
error in their ensemble average is much smaller. The size of the clusters
range from $10^3$ to $10^6$, and for each given size an ensemble of 1000 was
taken.

In Figure~\ref{fig-crossover} we show the penetration depth and compare its
crossover to that of the squared absolute value of the Laurent coefficients.
All of them seem to have asymptotic scaling of the form: 
\begin{equation}
r_o^2 (\alpha + \beta n^{-0.3}).
\end{equation}
The numerical results obtained by the HL method are close, especially for
larger cluster sizes, to those shown on Fig.~\ref{fig-crossover}.

What we have shown in this paper is that we can unify many of the
puzzling results on slow crossovers in DLA. The conformal map approach
gave us a strong indication that the crossovers associated with the
low order Laurent coefficients are shared by many quantities, but that
the asymptotic behavior is that all the characteristic length scales of
the growth will scale as $N^{1/D}$. 

In fact, this behavior is even more general than it appears. There is
another characteristic quantity with units of squared length, namely the
ensemble fluctuation of the squared radius \cite{Benny} : 
\begin{equation}
\delta (r_o^2) \equiv [<r_o^4> - <r_o^2>^2]^{1/2}
\label{sharpening}
\end{equation}
where $<>$ denotes an ensemble average.
The significance of $\delta (r_o^2)$ is that it is the spread, for
different clusters in the ensemble, of $r_o^2 N^{-2/D}$.  This is not
the same sort of object as those that we have been discussing, which
are averaged properties of individual clusters.  As was discussed in
Ref. \cite{Benny}, $\delta (r_o^2)$ appears to scale more slowly than
$r_o^2$.  Note, however, that Ref. \cite{Benny} used the HL method, so
that only small $N$ were available.

\begin{figure}
    \epsfxsize=3.4in
    \noindent\hfil\epsfbox{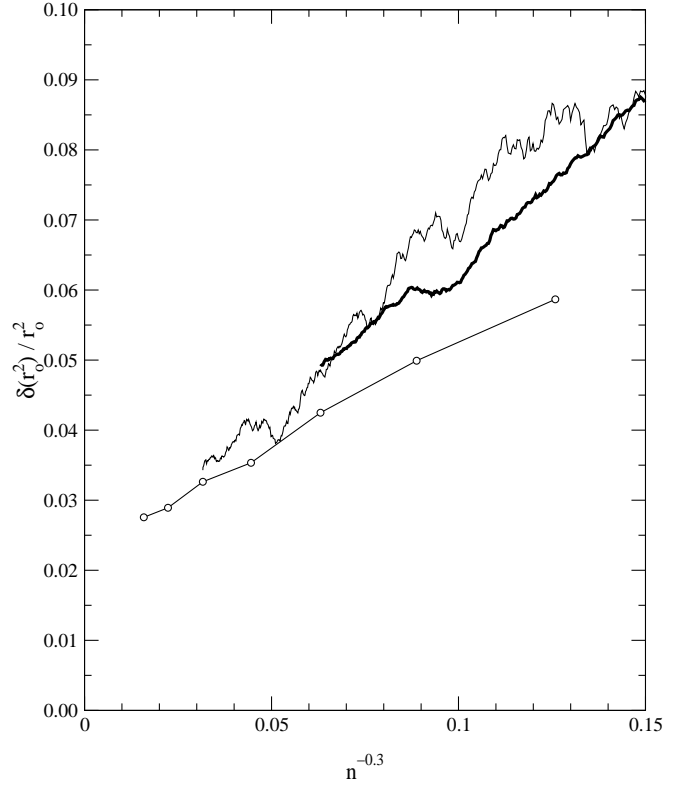}\hfil\bigskip\bigskip
    \caption{The relative distribution width of the cluster radius:
    $\delta(r_o^2)/r_o^2$. The distribution width $\delta(r_o^2)$ obeys the
    same correction to scaling as the other quantities of squared length
    dimension,
    see Fig.~\ref{fig-crossover}. The circles represent conventionally
    grown DLAs, average over 1000 clusters up to size $10^6$; the
    continuous lines are produced by the HL method: 400 clusters up to size
    $10^4$ (thick line) and 30 clusters up to size $10^5$ (thin line).}
    \label{fig-sharpening}
\end{figure}

If we go to large $N$ we find that $\delta (r_o^2)$ acts in the
same way as $\xi_\parallel^2$, with the same correction to
scaling: The apparent ensemble sharpening of the radius is also
a crossover.
This is shown in Figure~\ref{fig-sharpening}. Once more, directly
generating the charge distribution with random walkers allowed us to go to
large $N$ and reveal the crossover, which was not evident in the HL method.

These results, and the last one in particular, give rise to the
suspicion that all of the slowly scaling quantities 
in DLA growth are slaved to some
underlying variable. Such a view was proposed some time ago by Barker
and Ball \cite{BB}. In a future publication we will elaborate on this
idea in view of the present understanding of DLA \cite{BSS}.
We think that the considerations in this paper can be extended to 
give rise to a very detailed understanding of DLA clusters.

LMS and ES are supported by DOE grant DEFG-02-95ER-45546.
Part of this work was carried out at the Isaac Newton Institute,
whose support under the programme 'Mathematics and Applications of
Fractals' is gratefully acknowledged (RCB and LMS).

\begin{references}
\bibitem{excuse} This estimate differs from the calculation in Appendix~B 
of Ref.~\cite{Benny} because in that paper $|Z|$ in our Eq.\ref{estimate}
was implicitly replaced by $r_o$. We believe this to be too crude an
approximation.
\end{references}
\end{document}